# Epsilons Near Zero limits in the Mie scattering theory

M. Tagviashvili[*]

*Andronikashvili Institute of Physics, 6 Tamarashvili st, 0177, Tbilsi, Georgia*
[*] *madona.tagviashvili@grt.ge*

**Abstract:** The classical Mie theory - electromagnetic radiation scattering by the homogeneous spherical particles - is considered in the epsilon near zero limits separately for the materials of the particles and the surrounding medium. The maxima of a scattered transverse electrical (TE) field for the surrounding medium materials with the epsilon near zero limits are revealed. The effective multipole polarizabilities of the corresponding scattering particles are investigated. The possibility to achieve magnetic dipole resonance and accordingly to construct metamaterials with negative refractive index for the aggregates spherical particles in surrounding medium with the epsilon near zero limits is considered.

The scattering, absorption and extinction of the plane electromagnetic wave by spherical homogeneous particles was initially developed by Gustav Mie at 1905 and has a great number of applications in many fields of science and engineering. The Mie theory was intensively elaborated during the last century described more generalize scattering process with different kind of the scattering particles and the surrounding medium. In this article, the dielectric properties of the scattering particles and surrounding medium bulk materials are considered in the epsilon near zero limits $\varepsilon \to 0$. The peculiarities of epsilon near zero materials [1] revealed new features and singularity in Mie theory also.

As was shown by classical Mie theory, the optical behaviour of the system particles and its surrounding medium can be described by ratio $x = \frac{2\pi\delta}{\lambda_0}$ and by the refractive indices of the particles - $n_\alpha = \sqrt{\varepsilon_\alpha \mu_\alpha}$ (indexed by $\alpha$) and of the surrounding medium $n_\beta = \sqrt{\varepsilon_\beta \mu_\beta}$ (indexed by $\beta$), where $\delta$ is the particles radius, $\lambda_0$-the incident electromagnetic radiation's wavelength in vacuum, $\varepsilon_\alpha, \mu_\alpha$ and $\varepsilon_\beta, \mu_\beta$ are the dielectric permittivity and magnetic permeability of the materials inside particles and of the surrounding medium consequently. The incident, scattered and passed waves are expressed by the expansion in series - modes of the orthogonal spherical functions. The Mie coefficients are the weights of the expansion series. They are derived from continuity conditions of electrical and magnetic fields' tangential components at the spherical surface of the particles. Particularly, the scattered fields are described by Mie scattering coefficients $a_n$ and $b_n$, where coefficients $a_n$ apply to the transverse magnetic (TM) scattered fields and coefficients $b_n$ apply to the transverse electric (TE) scattered fields.

On the other hand, it is well known [2], that scattered electromagnetic field of an isolated sphere is equivalent to (that of ) the coherent ensemble of ideal point multipoles with appropriately chosen effective multipole polarizabilities. Each mode of the scattered electromagnetic field could be represented as independent collective response driven by the electric and magnetic fields of the corresponding multipole amplitudes in the orthogonal expansion driving wave. It can be shown that each of effective multipole polarizabilities are proportional to the corresponding partial wave amplitudes of the scattered wave expressed via the Mie coefficients $a_n$ and $b_n$. In particular the effective electric dipole polarizability of an isolated sphere $\alpha_e$ can be expressed via the TM field's first-dipole modes Mie scattering coefficient $a_1$:

$$a_e = i\frac{6\pi}{k_\beta^3} a_1 \qquad (1)$$

The magnetic dipole polarizability of an isolated sphere $\alpha_\mu$ can be found from the TE field's first - dipole modes Mie coefficient $b_1$:

$$a_\mu = i\frac{6\pi}{k_\beta^3}b_1 \qquad (2)$$

The scattering coefficients $a_n$ and $b_n$ denominator's zeros corresponds to the electric multipole and magnetic multipole resonances correspondingly.

Ordinarily [3], [4], [5], [6] the surrounding medium is considered to be passive and non absorptive. The system is described by the relative refractive index $n = \frac{k_\alpha}{k_\beta} = \frac{n_\alpha}{n_\beta}$ and by the ratio of particle's radius to the radiation's wavelength $\lambda_\beta$ in the surrounding medium $z_\beta = k_\beta \delta = \frac{2\pi\delta}{\lambda_\beta} = n_\beta x$. Accordingly the coefficients $a_n$ and $b_n$ are commonly expressed as functions of the variables $z_\beta$ and $nz_\beta$:

$$a_n = \frac{\mu_\beta n \psi_n(nz_\beta)\psi_n'(z_\beta) - \mu_\alpha \psi_n(z_\beta)\psi_n'(nz_\beta)}{\mu_\beta n \psi_n(nz_\beta)\zeta_n'(z_\beta) - \mu_\alpha \zeta_n(z_\beta)\psi_n'(nz_\beta)} \qquad (3)$$

$$b_n = \frac{\mu_\beta \psi_n(nz_\beta)\psi_n'(z_\beta) - \mu_\alpha n \psi_n(z_\beta)\psi_n'(nz_\beta)}{\mu_\beta \psi_n(nz_\beta)\zeta_n'(z_\beta) - \mu_\alpha n \zeta_n(z_\beta)\psi_n'(nz_\beta)} \qquad (4)$$

Where $\psi_n(y) = \sqrt{\frac{\pi y}{2}} J_{n+1/2}(y)$ - is the modified Bessel function and

$\zeta_n(y) = \sqrt{\frac{\pi y}{2}} H_{n+1/2}(y) = \sqrt{\frac{\pi y}{2}} (J_{n+1/2}(y) - iN_{n+1/2}(y))$ - is the modified Henkell function.

The expansion in series of the expressions (3) and (4) with small values of $z_\beta \to 0$ (the case for small particles when $\lambda_\beta >> 2\pi\delta$) gives well known Rayleigh approximation for the first mode's scattering coefficients:

$$a_1 = -i\frac{2}{3}(z_\beta)^3 \frac{\varepsilon_\alpha - \varepsilon_\beta}{\varepsilon_\alpha + 2\varepsilon_\beta} + O(z_\beta^5) \qquad (5)$$

$$b_1 = -i\frac{2}{3}(z_\beta)^3 \frac{\mu_\alpha - \mu_\beta}{\mu_\alpha + 2\mu_\beta} + O(z_\beta^5) \qquad (6)$$

The higher, n-th modes scattering $a_n$ and $b_n$ coefficients in the Rayleigh approximation has the smallness of the order equal to $z_\beta^{2n+1}$:

In the case of nonmagnetic particles $\mu_\alpha = 1$ and surrounding medium with $\mu_\beta = 1$, the coefficient $b_n$ and accordingly the magnetic dipole polarizability is $z_\beta^2$ times less then coefficient $a_n$ and accordingly the electric dipole polarizability. No magnetic dipole resonance exists for small particles $z_\beta = \frac{2\pi\delta}{\lambda_\beta} \to 0$.

The novelty of this work consist in consideration of the expression (3) and (4) as functions of the independent variables $z_\alpha = \frac{2\pi\delta}{\lambda_\alpha} = n_\alpha x$ and $z_\beta = \frac{2\pi\delta}{\lambda_\beta} = n_\beta x$, instead of the variables $z_\beta$ and $n = \frac{n_\alpha}{n_\beta}$ That means to investigate the dependence of the scattered field on the dielectric and magnetic properties of the surrounding medium and the particles separately.

Multiplying the (3) and (4) expressions' numerator and denominators by the $n_\beta x$ we receive the new expressions for the coefficients $a_n$ and $b_n$, were each of the spherical functions have arguments as one of the two independent variables $z_\alpha$ and $z_\beta$ [8]:

$$a_n = \frac{\mu_\beta z_\alpha \psi_n(z_\alpha)\psi_n'(z_\beta) - \mu_\alpha z_\beta \psi_n(z_\beta)\psi_n'(z_\alpha)}{\mu_\beta z_\alpha \psi_n(z_\alpha)\zeta_n'(z_\beta) - \mu_\alpha z_\beta \zeta_n(z_\beta)\psi_n'(z_\alpha)} \quad (7)$$

$$b_n = \frac{\mu_\beta z_\alpha \psi_n'(z_\alpha)\psi_n(z_\beta) - \mu_\alpha z_\beta \psi_n'(z_\beta)\psi_n(z_\alpha)}{\mu_\beta z_\alpha \psi_n'(z_\alpha)\zeta_n(z_\beta) - \mu_\alpha z_\beta \zeta_n'(z_\beta)\psi_n(z_\alpha)} \quad (8)$$

The expressions (9) and (10) can be examined separately relatively to the variables $z_\alpha$ and $z_\beta$. The first terms in $\psi_n(z)$ and $\zeta_n(z)$ functions' expansion series are:

$$\psi_n(z) = A_n z^{n+1} + O(z^{n+3}) \quad (9)$$

$$\zeta_n(z) = -B_n \frac{i}{z^n} + O\left(\frac{1}{z^{n-1}}\right) \quad (10)$$

where coefficients $A_n$ and $B_n$ can be expressed as: $A_n = \frac{2^{n+1}(n+1)!}{(2n+2)!}$ and $B_n = \frac{(2n)!}{2^n n!}$

If the epsilon near zero limits characterizes the bulk materials inside the scattering particles as $z_\alpha \to 0$ and the surrounding medium is with an arbitrary and independent parameter $z_\beta$, the Mie coefficients $a_n$ and $b_n$ tend to the expansion procedures using $z_\alpha$ as the parameter of smallness:

$$a_n = \frac{\psi_n(z_\beta)}{\zeta_n(z_\beta)} + O(z_\alpha^2) \quad (11)$$

$$b_n = \frac{\mu_\alpha z_\beta \psi_n'(z_\beta) - (n+1)\mu_\beta \psi_n(z_\beta)}{\mu_\alpha z_\beta \zeta_n'(z_\beta) - (n+1)\mu_\beta \zeta_n(z_\beta)} + O(z_\alpha^2) \quad (12)$$

All first terms of coefficients $a_n$ and $b_n$ expansions are periodical functions of variable $z_\beta$ with the equal magnitudes and without any points of singularity. The scattering efficiency $Q_{sca}$ has expression [6]

$$Q_{sca} = (2/z_\beta^2) \sum_{n=1}^{\infty} (2n+1)\{|a_n|^2 + |b_n|^2\} \tag{13}$$

The calculation of the sum (15) is roughly limited by four modes. It was shown that scattering efficiency $Q_{sca}$ is a smooth function with the maximum about the point $z_\beta = \pi$ (Fig.1).

As Opposite to above, the scattering coefficients of the particles' aggregate in the surrounding medium with the epsilon near zero limit $z_\beta \to 0$ has certain singular points at certain $z_\alpha$ values. In this limit, the incident radiation's wavelength in the medium greatly exceeds particles' radius $\lambda_\beta \gg 2\pi\delta$. If we Consider the parameter $z_\alpha$ as an independent certain finite value, then the limit $z_\beta \to 0$ tends to the following expansions of the Mie coefficients $a_n$ and $b_n$:

$$a_n = -i \frac{A_n}{B_n} \frac{(n+1)}{n} (z_\beta)^{2n+1} + O(z^{2n+2}) \tag{14}$$

$$b_n = i \frac{A_n}{B_n} \frac{(n+1)\mu_\alpha \psi_n(z_\alpha) - \mu_\beta z_\alpha \psi_n'(z_\alpha)}{n\mu_\alpha \psi_n(z_\alpha) + \mu_\beta z_\alpha \psi_n'(z_\alpha)} (z_\beta)^{2n+1} + O(z_\beta^{2n+2}) \tag{15}$$

Mie coefficients $a_n$ and $b_n$ of n-th modes are $z_\beta^2$ times less then $a_{n-1}$ and $b_{n-1}$ - the coefficients of the $(n-1)$-th modes correspondingly. The dipole, quadruple, octuple, hexadecapole and higher multipole electric and magnetic polarizabilities are proportional to $\approx \frac{a_n}{k_\beta^{2+n}}$, $\approx \frac{b_n}{k_\beta^{2+n}}$ accordingly. Each of them has the $k_\beta^{n-1}$ order of smallness. The coefficients $b_n$ have their singular points. The TE field's resonances appear when $z_\alpha$ parameter is the solution of equation (18)

$$n\mu_\alpha \psi_n(z_\alpha) + \mu_\beta z_\alpha \psi_n'(z_\alpha) = 0. \tag{16}$$

Fig. 2 presents Mie coefficient $a_n$ and $b_n$ in a logarithmic coordinates versus to particle's parameter $z_\alpha$ for given small $z_\beta = 0.01 + i*0.00001$.

If the particles and surrounding medium are nonmagnetic $\mu_\alpha = 1$, and $\mu_\beta = 1$, the first mode's Mie first mode's coefficient can be expressed as :

$$a_1 = -i \frac{2}{3} z_\beta^3 \tag{17}$$

$$b_1 = i \frac{2}{3} \frac{(3 - z_\alpha^2)\sin(z_\alpha) - 3z_\alpha \cos(z_\alpha)}{z_\alpha^2 \sin(z_\alpha)} z_\beta^3 \tag{18}$$

Despite of Rayleigh approximation in case of the epsilon near zero limit for the surrounding medium, the effective electric and magnetic polarizabilities in the nonmagnetic medium are commonly characterised with the equal-$z_\beta^3$ smallness of order. The effective magnetic polarizabilities' peaks of the maxima appare are at the points $z_\alpha$ satisfying the equation $\sin(z_\alpha) = 0$.

Fig. 3 demonstrates the scattering efficiency $Q_{sca}$ as a function of $z_\alpha$ and $z_\beta$. The sum is reasonably limited by four modes. The scattering efficiency peaks represent the resonances of the TE scattered field's first modes at $z_\alpha = n_\alpha \frac{2\pi\delta}{\lambda_o} = \pi m$, $m = 1,2,3,....$ .

In other words, we have evidence of the existence of the effective magnetic dipole resonances in the medium, where a dipole approximation $\lambda_\beta >> 2\pi\delta$ is valid. The Clausius-Mossotti relations (20) and (21) which describe the effective permittivity $\varepsilon^{eff}$ and an effective permeability $\mu^{eff}$ of the media with implemented particles have also reasonable validities [7].

$$\varepsilon^{eff} = \varepsilon_\beta \frac{1 + 2\frac{Na_e}{3\varepsilon_\beta}}{1 - \frac{Na_e}{3\varepsilon_\beta}} \quad (19)$$

$$\mu^{eff} = \frac{1 + 2a_\mu N/3}{1 - a_\mu N/3} \quad (20)$$

The volume density of the particles in the material is denoted by $N = \frac{1}{d^3}$ and the filling function is given by $f = 4\pi N\delta^3/3 = 4\pi/3(\delta/d)^3$, where $d^3$ is a volume of the unit cell.

Using the expressions for the effective magnetic dipole and electric dipole polarizabilities (1) and (2), the Clausius-Mossotti relations can be expressed as:

$$\varepsilon^{eff} = \varepsilon_\beta \frac{1 + 2f}{1 - f} \quad (21)$$

$$\mu^{eff} = \frac{1 + 2f \frac{(3 - z_\alpha^2)\sin(z_\alpha) - 3z_\alpha \cos(z_\alpha)}{z_\alpha^2 \sin(z_\alpha)}}{1 - f \frac{(3 - z_\alpha^2)\sin(z_\alpha) - 3z_\alpha \cos(z_\alpha)}{z_\alpha^2 \sin(z_\alpha)}} \quad (22)$$

The effective permittivity and effective permeability simultaneously achieve the negative values $\varepsilon^{eff} \to 2\varepsilon_b$, $\mu^{eff} \to -2$, if the surrounding medium's bulk dielectric permittivity is $\varepsilon_\beta \to 0$ near zero and negative $\varepsilon_\beta < 0$ and a radius $\delta$ of identical particles are equal to $\frac{m}{2}\frac{\lambda_0}{n_\alpha}$, where $n_\alpha$ is particles' bulk refractive index, $m = 1,2,3,....$ and $\lambda_0$ is the wavelength in **a** vacuum. That means that it is reasonable to consider the negative effective refractive index for the aggregates of identical spherical particles implanted in the surrounding medium with an epsilon near zero.

The metals could be regarded as the most preferred alternatives of an epsilon near zero material. The dielectric property of the metals are described via Drude-Sommerfeld model:

$$\varepsilon(\omega) = \varepsilon_\infty - \frac{\omega_p^2}{\omega(\omega + i\gamma_0)} \qquad (23)$$

where $\gamma_0$ is the electron relaxation rate, $\varepsilon_\infty$-the band electrons contribution, $\omega_p$-bulk plasma frequency could be accepted from the references [8]. For example, the negative epsilon near zero medium for gold takes place, when the incident light wavelength is about 430 nm ($\omega_p = 9.1 eV$, $\varepsilon_\infty = 9.84$, $\gamma_0 = 0.035 nc^{-1}$). For silver materials, the negative epsilon near zero can be achieved, when the incident light wavelength is about 260 nm ($\omega_p = 9.0 eV$, $\varepsilon_\infty = 3.7$, $\gamma_0 = 0.025 nc^{-1}$). Corresponding magnetic dipole resonances occurs for the particles with radius $\delta = 215$ nm, if the particles are implanted in gold and filled with vacuum $n_\alpha = 1$. We have magnetic dipole resonances for the particles filled with non dispersive materials with $n_\alpha = 2$ implanted in the gold when the radius of the particles is about 107 nm. For the particles implanted in silver with an epsilon near zero permittivity magnetic dipole resonance occurs for the particles with radiuses 130 nm and 65 nm if the particles itself consist of materials with refractive indexes $n_\alpha = 1$ and $n_\alpha = 2$ consequently.

Fig. 4 represents the effective refractive index profile for silver medium with implanted spherical identical particles as a function of the particles radius $\delta$. The bulk refractive index of particles is $n_\alpha = 1$, the incident lights wavelength is equal $\lambda_0 = 262$ nm and the system is characterized with filling function $f = 0.01$.

Fig. 5 represents the effective negative index areas of the wavelength in vacuum $\lambda_0$ and of the particle radius $\delta$ for the silver surrounding medium and implanted identical dielectric spherical particles ($n_\alpha = 1$, the Fig 5.a) and ($n_\alpha = 2$, Fig 5b). Fig. 6 represents the effective negative index areas of the wavelength in vacuum $\lambda_0$ and the particle radius $\delta$ for gold surrounding medium and implanted aggregate of particles with bulk refractive index $n_\alpha = 1$ (the Fig 6.a) and $n_\alpha = 2$ (the Fig 6.b).

The investigation of the Mie scattering for homogeneous (not layered) spherical nano-sized particles in the surrounding medium with epsilon near zero limits reveal the magnetic multipole resonances for s in the visible range of incident radiation. Similar to the other Mie resonances this phenomena can have applications in different fields of physics. Particularly, the investigation demonstrates possibility to achieve negative refractive index metamaterials constructed of the aggregates of spherical identical dielectric particles implanted in the metal surrounding medium in the visible range of the radiation.

The author is immensely grateful to Prof. V. Berezhiani for his comments and discussion.
The work was supported by ISTC under Grant No.G1366.

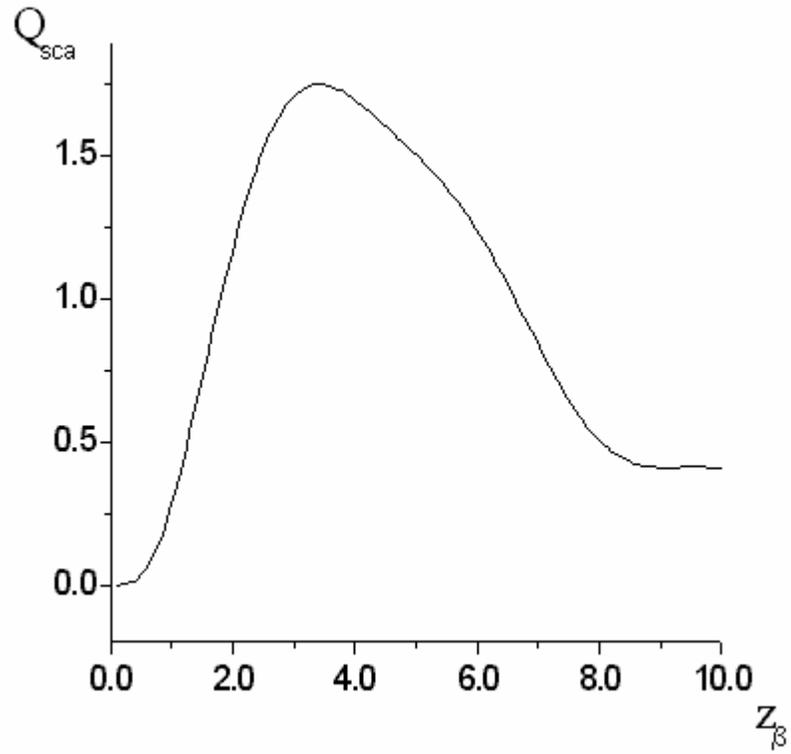

Fig.1 The scattering efficiency $Q_{sca}$ for the particles with epsilon near zero limit $z_\alpha = 0.01 + i*0.00001$ versus parameter $z_\beta$.

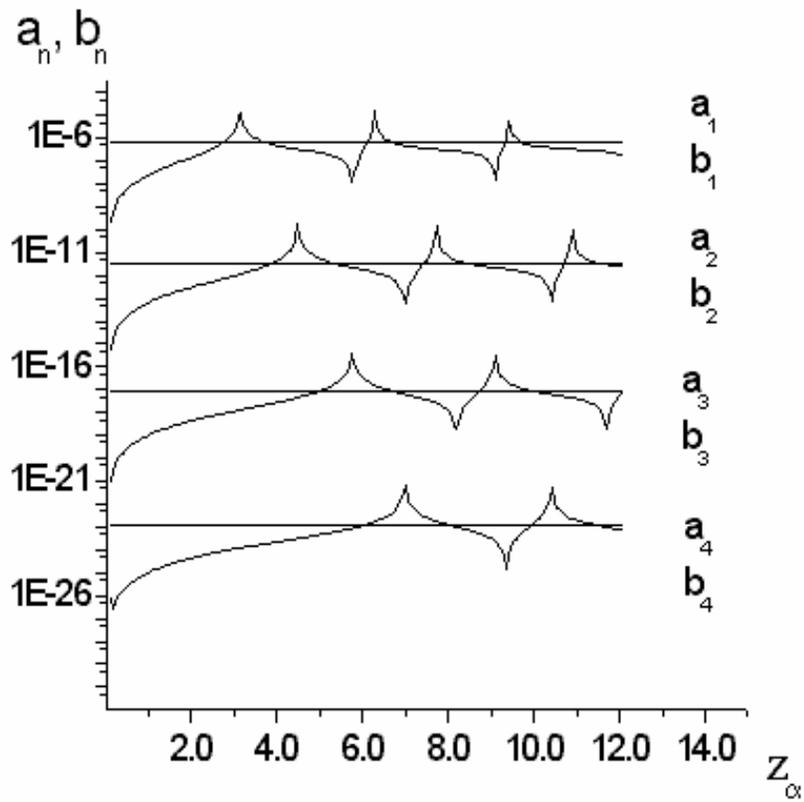

Fig .2 Mie coefficients $a_n$ and $b_n$ for the surrounding medium with epsilon near zero limit $z_\beta = 0.01 + i*0.00001$ as function of the parameter $z_\alpha$.

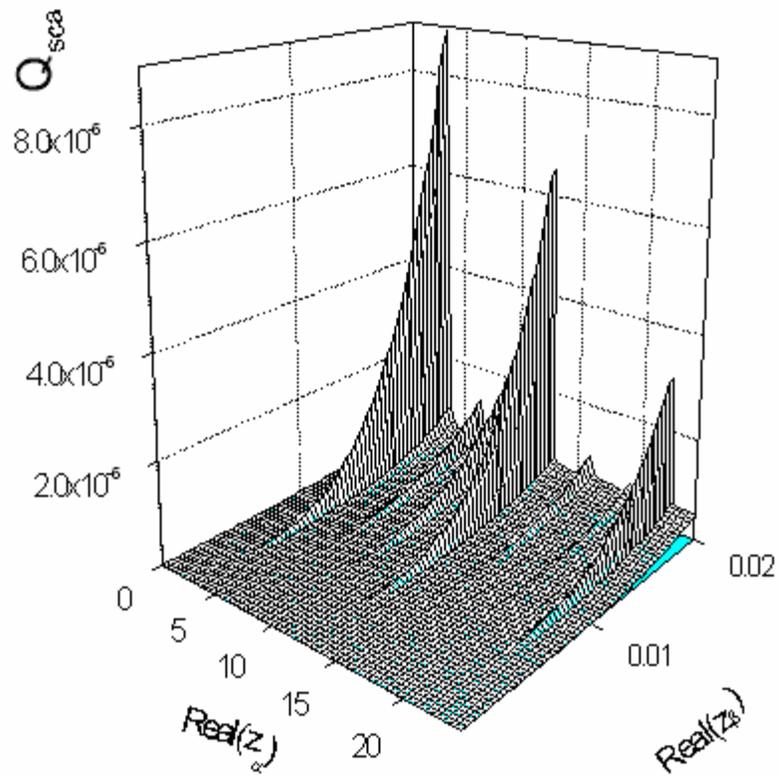

Fig.3. The scattering efficiency $Q_{sca}(z_\alpha, z_\beta)$.

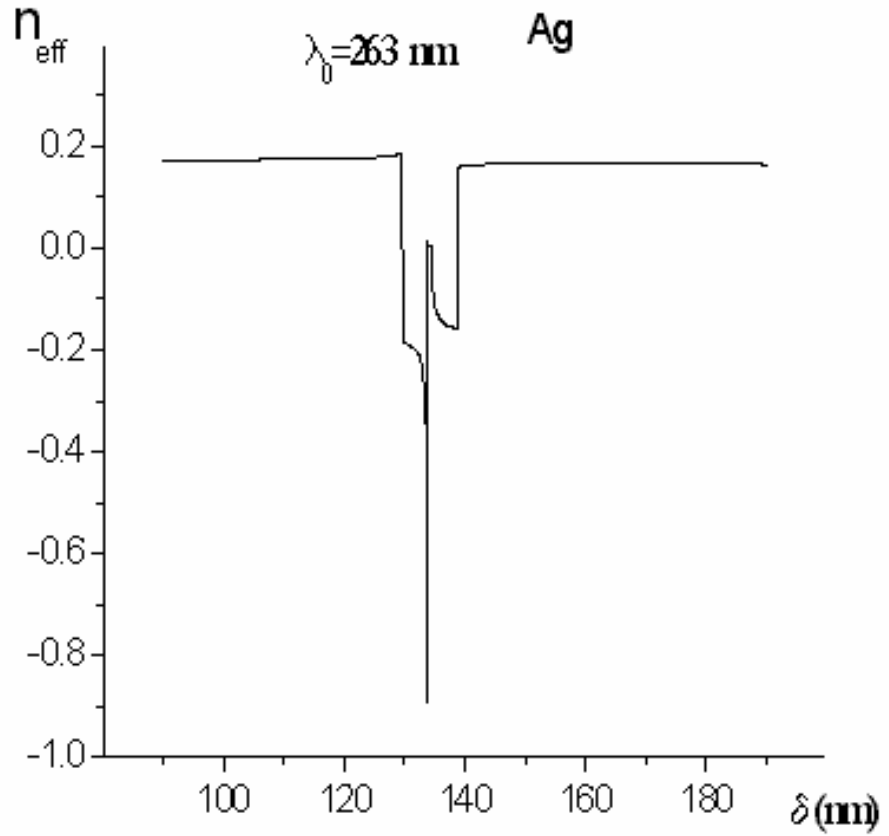

Fig.4. The effective refractive index of silver medium with implanted spherical identical particles versus the particles radius $\delta$. The particles' refractive index is equal to $n_\alpha = 1$, the incident lights wavelength - $\lambda_0 = 262$ nm and system's filling function - $f = 0.01$.

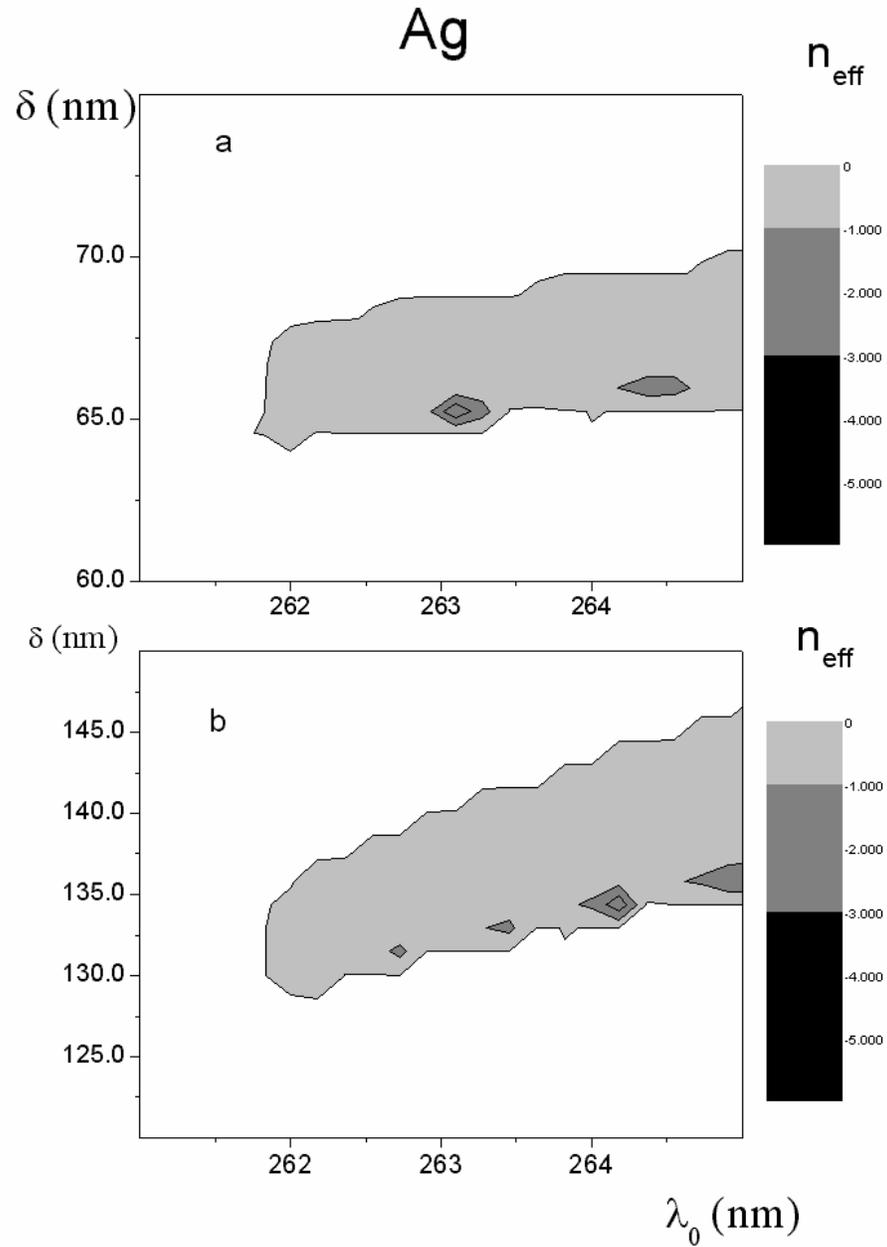

Fig. 5.The effective negative index areas of the wavelength in vacuum $\lambda_0$ and of the particle radius $\delta$ for the silver surrounding medium and implanted identical dielectric spherical particles. The particles' refractive indices are (a. $n_\alpha = 1$) and (b. $n_\alpha = 2$).

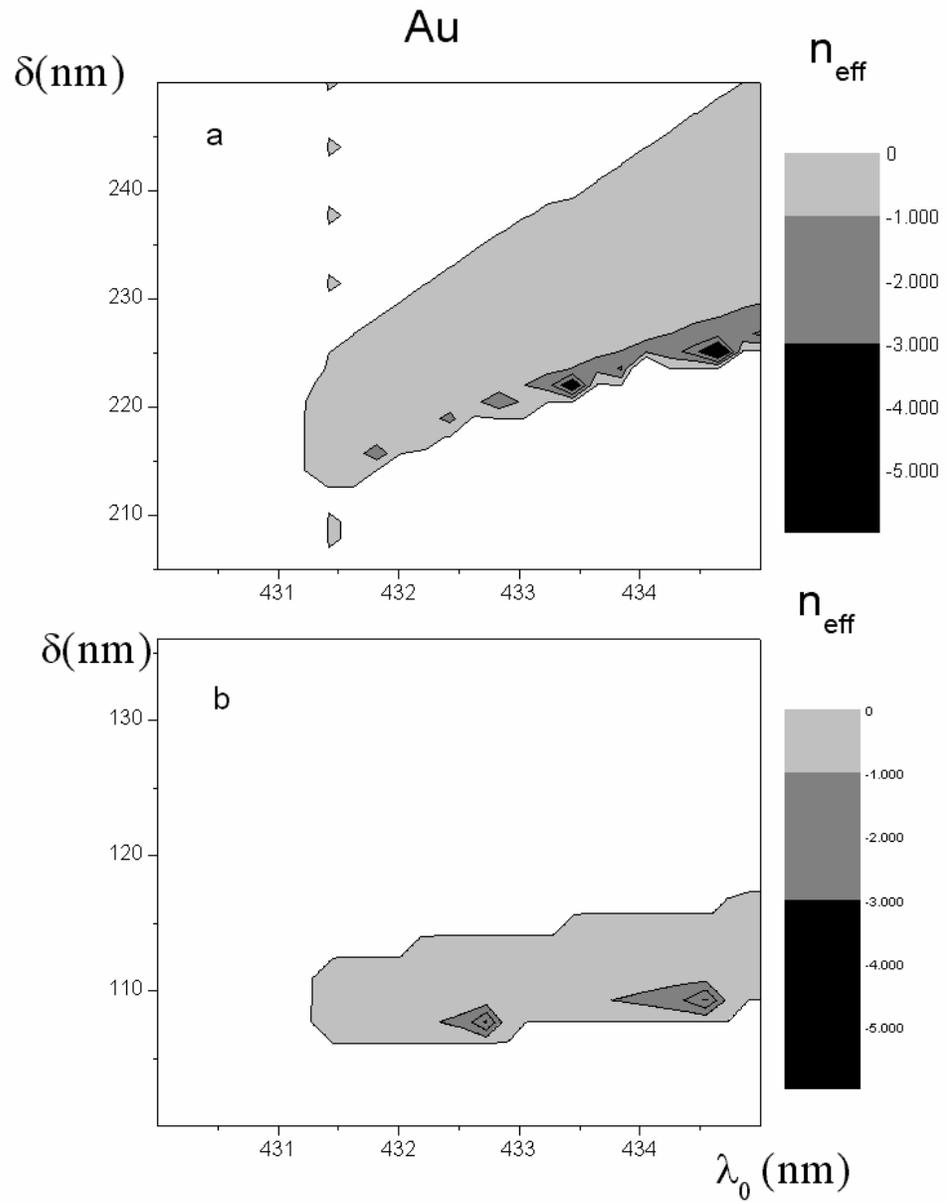

Fig. 6. The effective negative index areas of the wavelength in vacuum $\lambda_0$ and of the particle radius $\delta$ for the gold surrounding medium and implanted identical dielectric spherical particles (a. $n_\alpha = 1$) and (b. $n_\alpha = 2$).